\documentclass{aa}
\usepackage{txfonts}
\usepackage{natbib}
\bibpunct{(}{)}{;}{a}{}{,} 
\usepackage{graphicx}

\usepackage{natbib}
\usepackage[pdftex]{hyperref}
\hypersetup{colorlinks=true, pdfstartview=FitH, linkcolor={blue},citecolor={blue},urlcolor={red}}

\title{Mass limits for heavy neutrinos}
\author{Erik Elfgren
	\and Sverker Fredriksson}
\institute{Department of Physics, Lule\aa\ University of Technology, SE-971 87 Lule\aa, Sweden}
\offprints{Erik Elfgren,
\email{elf@ludd.ltu.se}}

\date{Received <date> / Accepted <date>}

\newcommand{\adsurl}[1]{\href{#1}{ADS}}
\newcommand{\elfurl}[1]{\href{#1}{LINK}}

\begin{document}

\abstract
{Neutrinos heavier than $M_Z/2\sim 45$ GeV are not excluded by particle physics data.
Stable neutrinos heavier than this might contribute to
the cosmic gamma ray background through annihilation in distant galaxies
as well as to the dark matter content of the universe.}
{ We calculate the evolution of the heavy neutrino density in the universe as a function
of its mass, $M_N$,
and then the subsequent gamma ray spectrum from annihilation of distant $N\bar{N}$ (from $0<z<5$).
}
{ The evolution of the heavy neutrino density in the universe is calculated numerically.
In order to obtain the enhancement
due to structure formation in the universe, we approximate
the distribution of $N$ to be proportional to that of dark matter in the GalICS model.
The calculated gamma ray spectrum is compared to the measured EGRET data.}
{
A conservative exclusion region for the heavy neutrino mass is 100 to 200~GeV,
both from EGRET data and our re-evalutation of the Kamiokande data.
The heavy neutrino contribution to dark matter is found to be at most 15\%.

}
{}

%

\maketitle



\section{Introduction}
The motivation for a fourth generation neutrino comes from the standard model of
particle physics. In fact, there is nothing in the standard model stating that there
should be exactly three generations of leptons (or of quarks for that matter). 


The present limits on the mass of a fourth generation of neutrinos are only
conclusive for $M_N\lesssim M_Z/2 \approx 46$ GeV \citep[p.~35]{2006JPhG...33....1Y}. This limit is obtained from the
measurement of the invisible width of the $Z^0$-peak in LEP,
which gives the number of light neutrino species, as $N_\nu = 2.9841 \pm 0.0083$ \citep{2001hep.ex...12021T}.

In 
\cite{2000PhLB..476..107M}, a fourth generation of fermions is found to
be possible for $M_{N}\sim 50$ GeV, while heavier fermions are shown to be unlikely. 
However, this constraint is only valid when there is a mixing between the generations
\citep{Novikov:2001md} and since this is not necessarily true, we will not take it for certain.

In the context of cosmology and astrophysics there are other contraints.
Light neutrinos, with $M_N\lesssim 1$ MeV, are relativistic when they decouple,
whereas heavier neutrinos are not. The light neutrinos must have $\sum m_\nu\lesssim 46$ eV
in order for $\Omega_\nu h^2<1$ to be valid \citep{2006PrPNP..57..309H}. For the dark matter (DM)
content calculated by \cite{2003ApJS..148..175S}, the bound is $\sum m_\nu\lesssim 12$~eV.
The number of light neutrino species are also constrained to $N_\nu = 4.2^{+1.2}_{-1.7}$ by
the cosmic microwave background (CMB), large scale structure (LSS),
and type Ia supernova (SNI-a) observations at 95\% confidence \citep{2006JCAP...01..001H}.

Neutrinos heavier than about 1 MeV, however, leave thermal equilibirum before
decoupling and therefore their number density drops dramatically,
see for example \cite{RevModPhys.53.1}. This will be discussed in more detail
in Sect.~\ref{se:Evolution}.

The most important astrophysical bound on heavy neutrinos comes from Kamiokande
\citep{1992PhLB..289..463M} and this will be considered separately in the end.

In
\cite{PhysRevD.52.1828}, it is found that the mass range $60\lesssim M_N \lesssim 115$ GeV
is excluded by heavy neutrino annihilation in the galactic halo.
However, according to \citet[p.~57]{2002PhR...370..333D} this constraint is based on
an exaggerated value of the density enhancement in our galaxy.

Other works constraining the heavy neutrino mass include
\cite{1998JETPL..68..685F,1999astro.ph..2327F} and
\cite{2004hep.ph...11093B}. There has also been a study of the gamma ray spectrum
of dark matter (DM) in general \citep{2007PhRvD..75f3519A}.

For an exhaustive review of modern neutrino cosmology, including current constraints
on heavy neutrinos, see \citet{2002PhR...370..333D}. It is concluded that
there are no convincing limits on neutrinos in the mass range $50\lesssim M_N \lesssim 1000$ GeV.
A review of some cosmological implications of neutrino masses and mixing angles can be found in
\cite{2003nema.conf...53K}.


In this paper we consider a stable fourth generation heavy neatrino with mass $M_N \gtrsim 50$ GeV
possessing the standard weak interaction. We assume that other particles of a fourth
generation are heavier and thus do not influence the calculations.

We assume a $\Lambda$CDM universe with
$\Omega_{tot} = \Omega_m + \Omega_\Lambda = 1$, where
$\Omega_m = \Omega_b + \Omega_{DM} = 0.135/h^2$, $\Omega_b = 0.0226/h^2$ and
$h = 0.71$ \citep{2003ApJS..148..175S},
using WMAP data in combination with other CMB datasets and large-scale structure observations 
(2dFGRS + Lyman $\alpha$). 

Throughout the article we use natural units, such that the speed of light, Planck's reduced constant
and Boltzmann's constant equal unity,
$c = \hbar = k_B = 1$.

If heavy neutrinos ($M_N\gtrsim 50$ GeV) exist, they were created in the early universe.
They were in thermal equilibrium in the early stages of the hot big bang, but froze out
relatively early.
After freeze-out, the annihilation of $N\bar N$ continued at an ever
decreasing rate until today.
Since those photons that were produced before the decoupling of photons
are lost in the CMB, only the subsequent $N\bar N$ annihilations contribute to the photon background
as measured on earth. 

The intensity of the photons from $N\bar N$-annihilation
is affected by the number density of heavy neutrinos, $n_N$,
whose mean density decreases as $R^{-3}$, where $R$ is the
expansion factor of the universe. However, in structures such as
galaxies the mean density will not change dramatically, and
since the number of such structures are growing with time, this
will compensate for the lower mean density.
Note that the photons are also
redshifted with a factor $R$ due to their passage through space-time.
This also means that the closer annihilations will give photons
with higher energy than the farther ones.

\section{Evolution of neutrino density} \label{se:Evolution}

Let us recapitulate the results of \cite{RevModPhys.53.1}.

The cosmic evolution of the number density, $n_X$, of a particle $X$ can, in general,
be written as
\begin{equation}\label{eq:ndot}
	\dot n_X = -n_X^2 \left<\sigma v\right> - 3H(t)n_X + \psi(t),
\end{equation}
where
$\left<\sigma v\right>$ is the
thermally averaged product of the mean velocity and total annihilation cross section for the particle, and
$H(t)=\dot R/R$ is the Hubble constant.
The term $-3H(t)n_X$ represents
the expansion of the universe, and
the production term is
$\psi(t) = n_{Xeq}^2 \left<\sigma v\right>$, where $n_{Xeq}$ is the
equilibrium concentration of particle $X$.

If we write $r_X = n_X/n_\gamma$, Eq. (\ref{eq:ndot}) can be expressed as
\begin{equation} \label{eq:rate}
	\dot r_X = -\left< \sigma v\right> n_\gamma(r_X^2 - r_{Xeq}^2),
\end{equation}
where
\begin{equation}
	r_{Xeq} \approx \left\{ \begin{array}{ll}
		1 & \textrm{if $\theta \equiv T/m_X>1$}\\
		\frac{(2\pi)^{-3/2}}{2\cdot \zeta(3)/\pi^2} g_s \theta^{-3/2}e^{-1/\theta} & \textrm{if $\theta<1$}.
		\end{array} \right.
\label{eq:rXeq}
\end{equation}
Here $\zeta(3)\approx 1.2020569$ is the Riemann zeta function, $T$ is the temperature, $m_X$ is the mass of particle $X$ and $g_s$ is the
number of spin states. For photons and electrons, $g_s=2$, while for massless
left-handed neutrinos, $g_s=1$. For reference, $\frac{(2\pi)^{-3/2}}{2\cdot \zeta(3)/\pi^2} \approx \frac 14$.

The value of the relative equilibrium concentration, $r_{Xeq}$, is derived from
\begin{equation}
	r_{Xeq} \equiv n_\gamma^{-1} n_{eq} = \frac{1}{2T^3\zeta(3)/\pi^2} \cdot \frac{1}{(2\pi)^3} \int \frac{4\pi p^2dp }{e^{E/T}+1},
\end{equation}
where the expressions for $n_\gamma$ and $n_{eq}$ were taken from \citet[Eq. 30]{2002PhR...370..333D}.

According to \citet[Eq 2.9]{RevModPhys.53.1}, freeze-out (equilibrium destruction) occurs when
the rate of change of the equilibrium concentration due to the temperature decrease is
higher than the reaction rates, which means that $2\left<\sigma v\right> n_\gamma r_{Xeq} t T / m~>~1$.
Until freeze-out, the relative particle density follows the equilibrium
density closely: $r_{fX} \approx r_{Xeq}$.
Hence, the relative density at the moment of freeze-out is
\begin{equation}
	r_{fX} = (2\left<\sigma v\right> n_\gamma t_f\theta_f)^{-1} \approx r_{Xeq},
\label{eq:rfX}
\end{equation}
where $t_f$ and $\theta_f=T_f/m_X$ are the time and relative temperature at freeze-out.

As the temperature decreases, the production term $r_{Xeq}$ will drop exponentionally,
such that the relic concentration of $X$ will be more or less independent of $r_{Xeq}$.
With this approximation ($r_{Xeq} = 0$), Eq.~\ref{eq:rate} can be solved for $t\rightarrow \infty$:
\begin{equation} \label{eq:relic}
	r_{0X} \approx \frac{1}{2\left<\sigma v\right> n_\gamma t_f\cdot(1+\theta_f)}
	 = \frac{1}{2\left<\sigma v\right>_f n_{\gamma f} \frac{3.68\cdot 10^{18}}{\sqrt{g_*(T_f)}} T_f^{-2}(1+\theta_f)},
\end{equation}
where we have used $tT^2 \approx 3.677\times 10^{18}/\sqrt{g_*}$ \cite[Eq.~37]{2002PhR...370..333D},
with $g_*(T_f)$ from
\citet[p.~65]{Kolb:EU90}
being the number of relativistic species in thermal
contact with the photons. Furthermore, $n_\gamma(t_0) = 2T_0^3\zeta(3)/\pi^2 \approx 0.24 T_0^3$ is the photon density today,
where the photon temperature today is $T_0=2.725$~K \citep{1999ApJ...512..511M}.
According to the standard model of particle physics, $g_*=106.75$ for $T\gtrsim 100$ GeV
($g_* \approx 200$ for supersymmetry models at yet higher temperatures).
If we assume that $\theta_f \ll 1$ (which we will later show to be reasonable),
we obtain $r_{0X} \approx r_{fX} \theta_f$,
which differs by a factor two from the result of \citet[Eq.~2.11]{RevModPhys.53.1}.
This is natural if they consider the density of $n_{N+\bar N}$ since our $r_{0X}$ is valid for $N$ and $\bar N$
separately.

In order to take into account the increase in temperature due to entropy conservation
after freeze-out of particle $X$, we must take
\begin{equation}\label{eq:n0}
	\left(\frac{n_{0X}}{\textrm{m}^{-3}}\right) = r_{0X}\frac{43/11}{g_{*S}(T_f)} n_\gamma(t_0) \approx \frac{6.88\times 10^{-57}}{\left<\sigma v\right>_fT_f(1+T_f/m_X)\sqrt{g_{*f}}}.
\end{equation}
(In fact $g_{*f}^{-1/2}$ should be written $g_{*Sf}^{-1}\cdot g_{*f}^{1/2}$ but for $T_f>0.1$ GeV, $g_{*Sf}=g_{*f}$.)

\begin{figure}[here!]
        \resizebox{\hsize}{!}{\includegraphics{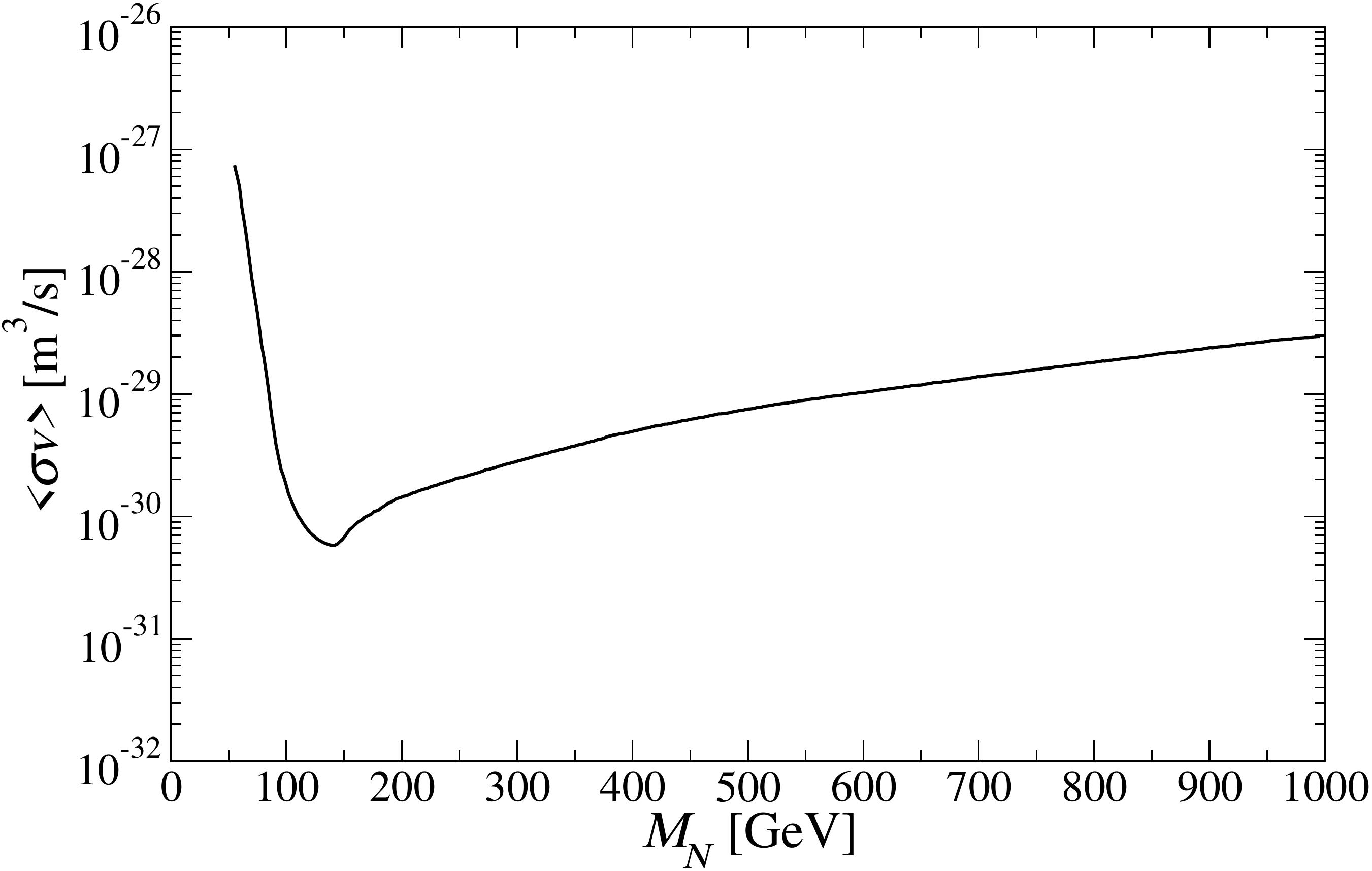}}
        \caption{The cross section times the velocity (in m$^3$/s) 
	of heavy neutrino annihilation $N\bar N$ as a function of their mass (in GeV) at freeze-out, $T=T_f$.}
\label{fig:sigmav}
\end{figure}

We now turn to the case of heavy neutrinos.
Since we wish to avoid the lenghty calculations
of the cross sections of heavy neutrinos \citep{1989NuPhB.317..647E}, we use \citet[Fig.~1 and Eq.~4]{PhysRevD.52.1828}
and solve
for $\left<\sigma v\right>$.
We assume that they use $g_*=g_*(T_f) \approx g_*(M_N/30)$, but the exact value does not change the result
in any significant way.
The resulting $\left<\sigma v\right>$ is presented in Fig.~\ref{fig:sigmav}.
The cross section drops from $M_N\sim 45$ GeV, where the $Z^0$
resonance peaks until the $W^+W^-$ annihilation channel starts to dominate at
$M_N \gtrsim 100$ GeV.

According to \cite{PhysRevD.52.1828}, 
the cross sections of heavy neutrinos can be estimated using the annihilation
channels
\begin{eqnarray}
	N\bar N &\rightarrow& Z^0 \rightarrow f\bar f \\
	N\bar N &\rightarrow& Z^0 \rightarrow W^+W^-.
\end{eqnarray}
There are several other possible annihilation channels for $ N\bar N \rightarrow W^+W^-$, like
$N\bar N \rightarrow L\bar L,\, H^0H^0,\, Z^0Z^0 \rightarrow W^+W^-$ and also interference between
$L$ and $Z^0$, as well as between $L$ and $H^0$. However, in the limit $s \rightarrow 4M_N^2$,
which is valid for cosmological heavy neutrinos,
the dominant channel is through s-channel $N\bar N \rightarrow Z^0$ \cite[p.~656]{1989NuPhB.317..647E}.
Furthermore, the other annihilation products, $N\bar N \rightarrow H^0H^0,\, Z^0Z^0$, are suppressed with
respect to $W^+W^-$-production \cite[p.~651,~656]{1989NuPhB.317..647E}. Hence, the above estimation
of the $\left<\sigma v\right>$ should be fairly accurate. If anything, it is slightly underestimated.

\begin{figure}[here!] 
        \resizebox{\hsize}{!}{\includegraphics{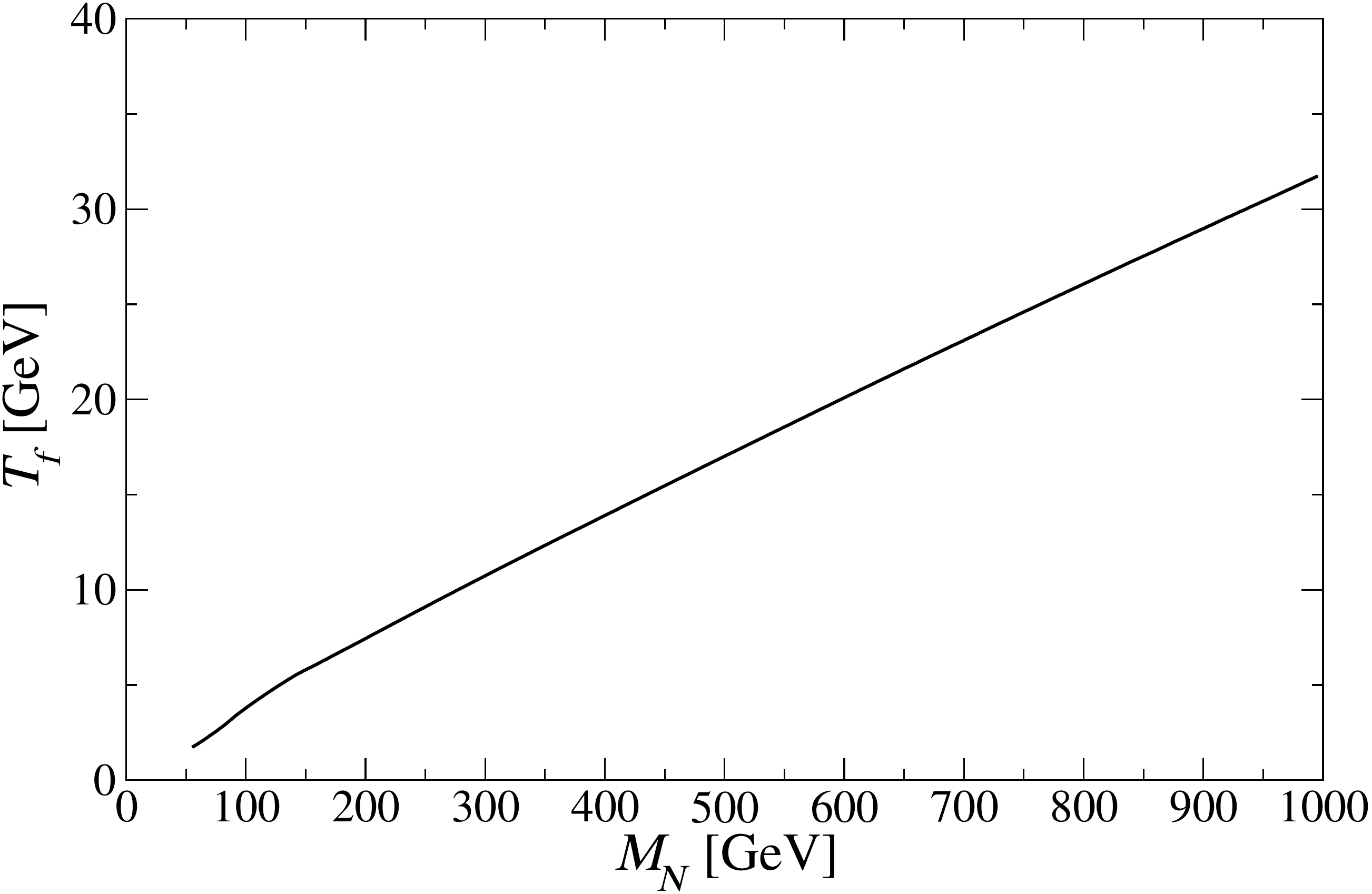}}
        \caption{The freeze-out temperature (in GeV$\approx 1.16\times 10^{13}$ K)
	of heavy neutrinos as a function of their mass (in GeV).}
\label{fig:Tf}
\end{figure}

Using Eqs.~\ref{eq:rfX} and \ref{eq:rXeq}, we can solve for $T_f=\theta_f\cdot M$.
The result is presented in Fig. \ref{fig:Tf}.
Note that although it looks like a straight line, it really is slightly curved.
We notice that $T_f/M_N \sim 1/30$, which shows our
assumption $M_N\gg T_f$ to be valid. This is also in agreement with previous
results, see e.g. \cite{Kolb:EU90}, where a value of $T_f/M_N \sim 1/20$ is quoted.

\begin{figure}[here!] 
        \resizebox{\hsize}{!}{\includegraphics{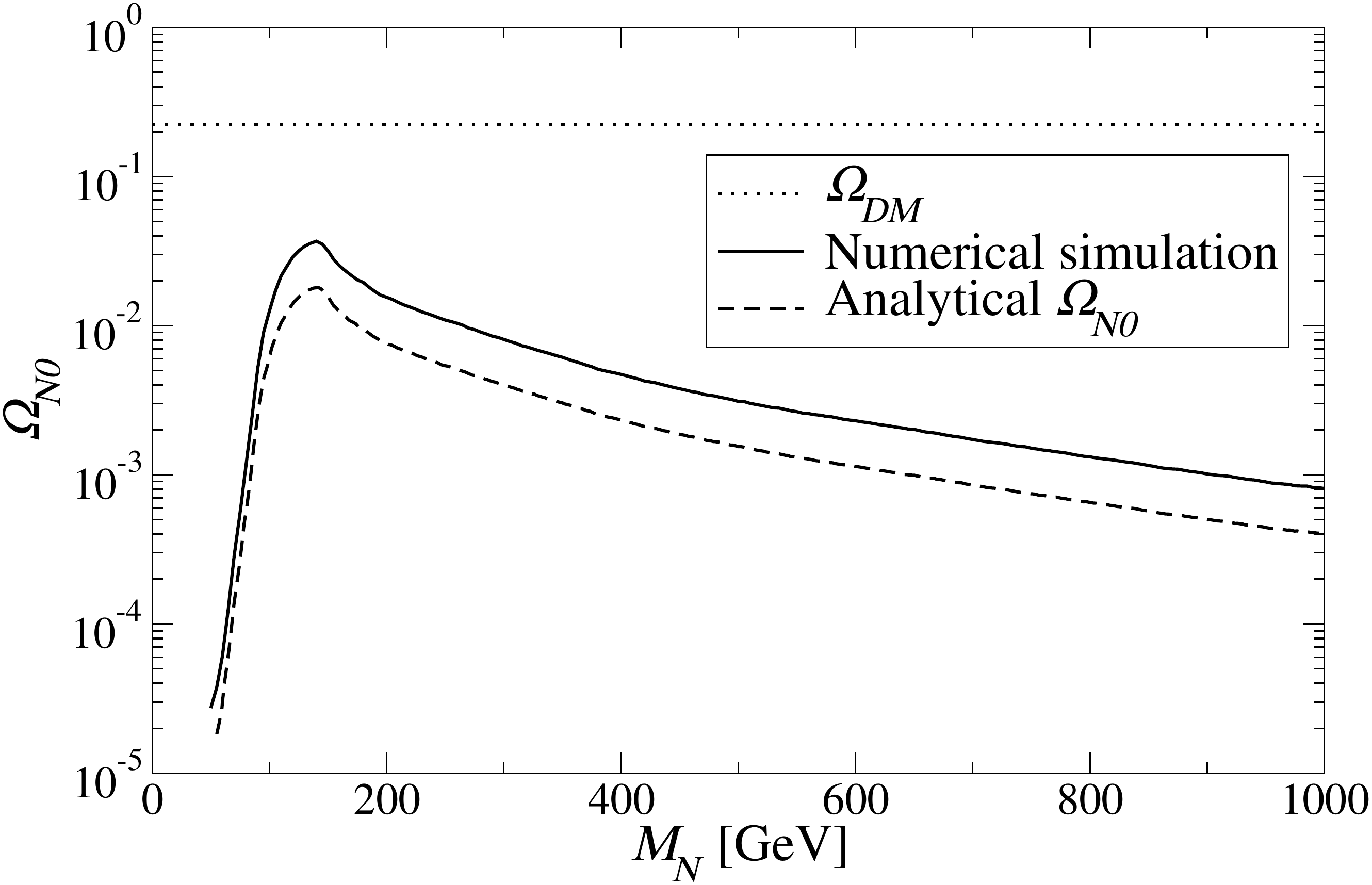}}
        \caption{ The relic relative density 
	of heavy neutrinos as a function of their mass (in GeV).}
\label{fig:Omega_N}
\end{figure}

We now return to Eq.~\ref{eq:n0} and apply it to the case of a heavy neutrino.
We plot the resulting relative relic neutrino density as a function of the mass $M_N$ in
Fig.~\ref{fig:Omega_N} using
$\Omega_N = 2M_N\cdot n_N(T_0)/\rho_c$, where $\rho_c\approx 9.47\times10^{-27}$ kg/m$^3$ is the critical density
of the universe.
The resulting heavy neutrino density is very similar to the one obtained by \citet[Fig.~1]{PhysRevD.52.1828}.
The numerical simulation also shown in the figure will be the subject of the next section.

\section{Numerical simulation of the neutrino density}
\label{sec:Num}
For comparison, we evaluate the evolution of the heavy neutrino density numerically.
Eq.~\ref{eq:ndot} can be rewritten in terms of the temperature, $T$:
\begin{equation}
	\frac{dn}{dT} = -\frac{dt}{dT} \left[ 3H(T)n(T) + \left<\sigma v(T)\right> \left(n(T)^2 - n_{eq}(T)^2\right)\right],
	\label{eq:Boltzmann}
\end{equation}
where
\begin{equation}
	n_{eq}(T) = r_{eq}n_\gamma = 2T^3\left(\frac{M_N}{2\pi T}\right)^{3/2} e^{-M_N/T}, \quad\quad (T<M_N)
\end{equation}
and the relation between time and temperature is given by
\begin{equation}
	\frac{dt}{dT} = \frac{-1}{H(T)}\left( \frac{1}{T} + \frac{dg_{*S}/dT}{3g_{*S}}\right),
\end{equation}
Here the Hubble constant is $H(T)=H_0\sqrt{\Omega(T)}$, where the total relative energy density of the universe is
\begin{equation}
	\Omega(T)= \Omega_R(T)\cdot R^{-4} + \Omega_M\cdot R^{-3} + \Omega_k\cdot R^{-2} + \Omega_\Lambda.
\end{equation}
The curvature term $\Omega_k = 0$ and the radiation density is
\begin{equation}
	\Omega_R(T) = \Omega_{R} \frac{g_*(T)}{g_*(T_0)}
\end{equation}
due to the reheating as particles freeze out.
The reheating also means that $R=g_{*S}^{-1/3}T_0/T$ \citep[p.~68]{Kolb:EU90}.
The number of relativistic species
still in thermal contact with the photons, $g_{*S}(T)$, is given in \citet[Fig.~1]{2003PhRvD..68b7702C}.
For the critical region $0.15<T<0.30$ GeV their Eqs.~8-9 have been used to calculate $dg_{*S}/dT$.
This updated value of $g_{*S}(T)$ is needed to evaluate $dg_{*S}/dT$ properly.

Using a fifth-order Runge-Kutta method with adaptive stepsize control, taken from Numerical Recipes
\cite[Ch.~16.2]{1992NumRec}, we solve for $n(T)$ in Eq.~\ref{eq:Boltzmann}  using 
the initial condition $n_i=n_{eq}(T_i=M_N/15)$, which is well within the region of thermal
equilibrium for the heavy neutrinos. The resulting
relative relic neutrino density is presented in Fig.~\ref{fig:Omega_N}, where
$\Omega_N = 2M_N\cdot n_N(T_0)/\rho_c$ as before.
We notice that the peak of the curve is $\Omega_N(M_N=140 \textrm{ GeV}) \approx 0.04$, which would
then account for $\sim$15\% of the dark matter content of the universe.

\begin{figure}[here!] 
        \resizebox{\hsize}{!}{\includegraphics{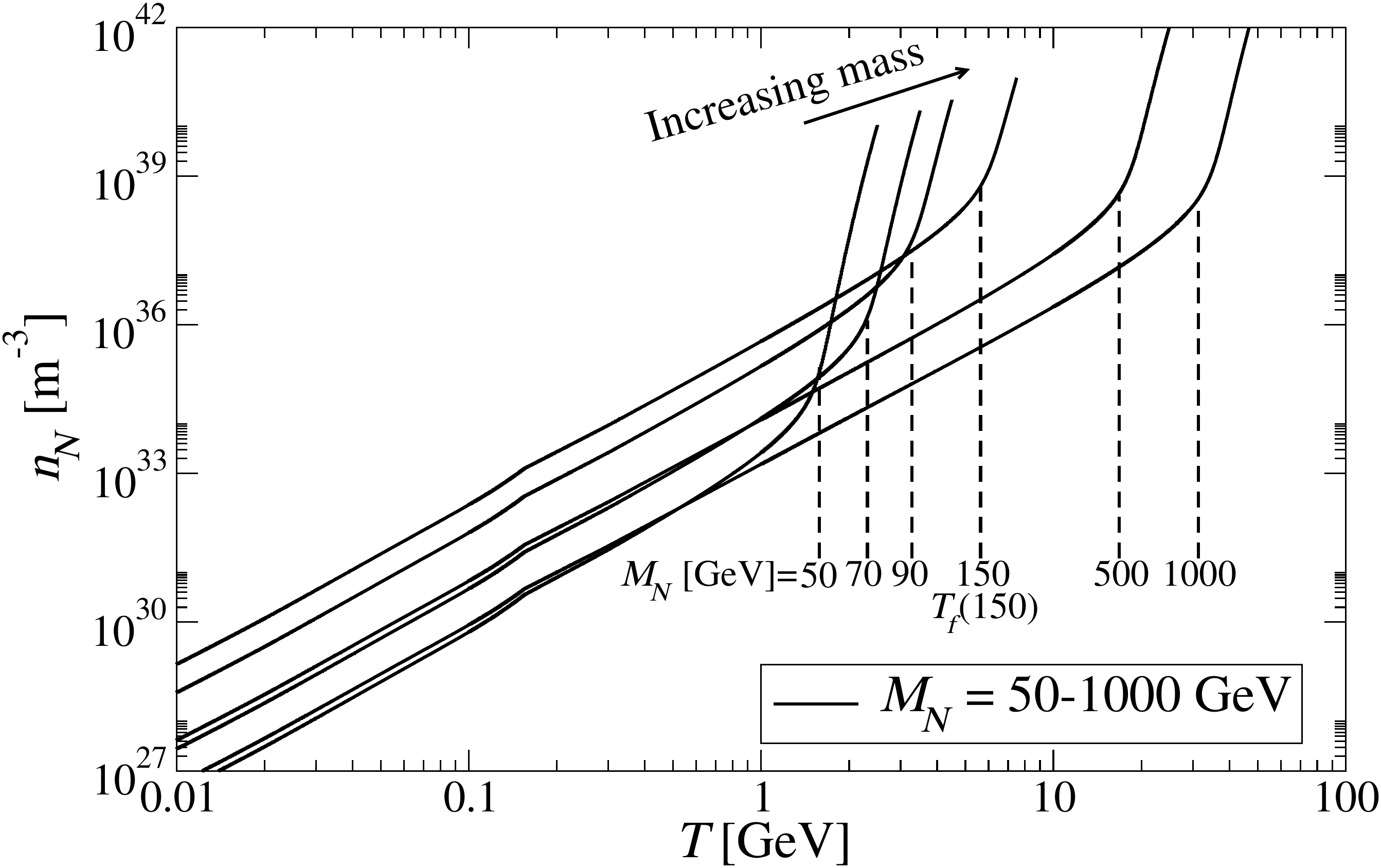}}
        \caption{ The number density 
	of heavy neutrinos (in m$^{-3}$) as a function of $T$ for masses 50, 70, 90, 150, 500 and 1000 GeV
	(increasing from left to right in the upper right corner).
	The dashed vertical lines represent the calculated value of $T_f$ in Fig.~\ref{fig:Tf}.
	Below $T=0.01$ GeV, the curves evolve as $(T/T_0)^3\cdot g_{*S}$.
}
\label{fig:n_T_N}
\end{figure}

For comparison, we plot the number density of heavy neutrinos (in m$^{-3}$) as a function of $T$
for masses 50, 70, 90, 150, 500 and 1000 GeV in Fig.~\ref{fig:n_T_N}. As we can see, the transition
between thermal equilibirum density and completely decoupled neutrino density is not sharp.
This is one of the reasons for the difference between the analytical and the numerical relative density
in Fig.~\ref{fig:Omega_N}.
Another reason for the difference is the inclusion of the change in $g_{*S}$ in the
evaluation of $dt/dT$.
The evolution of $g_{*S}(T)$ is the cause of the small "knee" in Fig.~\ref{fig:n_T_N} seen at $T\sim0.2$ GeV (the reheating from
the quark-hadron transition). Furthermore, when electrons fall out of thermal equilibirum at $T\sim 1$ MeV
there is another small knee, reducing again the heavy neutrino density somewhat.

\section{Dark matter simulations}
\label{sec:DM}
In Sect.~\ref{sec:Num}, we calculated the mean density of neutrinos in the
universe as a function of redshift and the mass of the heavy neutrinos. 
However, the neutrino annihilation rate, and thus the intensity from
their gamma spectrum, is proportional to the square of the neutrino density.
This means that inhomogeneities in the universe will tend to enhance the
gamma ray signal.

In this section we describe how we calculate the inhomogeneities as a
function of space and time, assuming only gravitational interaction
between the dark matter consisting of heavy neutrinos and other DM particles. 
The clumping factor (also known as the boost factor) can then be used to calculate the actual intensity
\begin{equation}
	\frac{dI}{dz} =  C(z)\frac{dI_0}{dz},
\end{equation}
where $dI_0/dz$ is the intensity contribution from redshift slice $dz$
for a homogeneous universe and $C(z)$ is the enhancement due to the clumping at redshift $z$.

The clumping factor has been calculated in different settings before, ranging from
\cite{2006PhRvD..73f3504B} for local clustering giving a clumping factor of $\sim5$
to \cite{2005Natur.433..389D} for mini-halos giving a clumping factor of
two orders of magnitude. 
For a discussion about the accuracy of approximating the enhancement with a single
clumping parameter, see \cite{2006astro.ph..3796L}, though they focus on antiprotons.

The spatial and temporal distribution of DM in the universe is calculated
with the GalICS program.
The cosmological N-body simulation that we are referring to throughout this paper is done
with the parallel tree-code developed by \cite{1999ninin}.
The initial mass power
spectrum is taken to be a scale-free ($n_s = 1$) one,
evolved as predicted by \cite{1986ApJ...304...15B} and normalized to
the present-day abundance of rich clusters with $\sigma_8$ = 0.88
\citep{1996MNRAS.282..263E}.
The DM density field was calculated from $z=35.59$ to $z=0$, giving 100
"snapshots", spaced logarithmically in the expansion factor.


The basic principle of the simulations is to distribute a number of
DM particles $N^3$ with mass $M_{{\rm DM}}$ in a box of size $L^3$.
Then, as time passes, the particles interact gravitationally, clumping together
and forming structures.
When there are at least 20 particles together, it is considered to be a DM halo.
It is supposed to be no other forces present than gravitation,
and the boundary conditions are assumed to be periodic.

In the GalICS simulations the side of the box used was
$L=100h^{-1}$ Mpc, and the number of particles was set to $256^3$, which implies a
particle mass of $\sim 5.51\times 10^{9}h^{-1} M_\odot$. Furthermore, for the
simulation of DM, the cosmological parameters were set to 
$\Omega_\Lambda = 2/3$, $\Omega_m = 1/3$ and $h=2/3$.
The simulations of the DM were done before the results from WMAP
were published, which
explains the difference between these parameters and the values 
used elsewhere in this paper, as stated in the introduction.
Nevertheless, the difference is only a couple of percent and should not seriously
alter the results. 

Between the initial halo formation at $z\sim 11$ and the current
epoch in the universe, there are 72 snapshots. In each snapshot
a friend-of-friend algorithm was used to identify virialized groups of at least
20 DM particles.
For high resolutions, it is clear that the mass resolution
is insufficient. Fortunately, the first 20-particle DM clump appears at $z=11.2$, while
the bulk of the clumping comes from $z\lesssim 5$, where the lack of resolution is 
no longer a problem.

In order to make a correct large-scale prediction of the distribution of the
DM, the size of the box would have to be of Hubble size, i.e., $\sim3000h^{-1}$ Mpc.
However, for a given simulation time, increasing the size of the box and
maintaining the same number of particles would mean that we lose in mass
resolution, which is not acceptable if we want to reproduce a fairly realistic
scenario for the evolution of the universe.

We will make the approximation that our single box, at different time-steps,
can represent the line of sight, and since we are only interested in the general
properties of the dark matter clumping, this approximation should be
acceptable.


\subsection{Validity of simulation}
GalICS is a hybrid model for hierarchical galaxy formation, combining
the outputs of large cosmological N-body simulations with simple, semi-analytic
recipes to describe the fate of the baryons within DM halos. The
simulations produce a detailed merging tree for the DM halos,
including complete knowledge of the statistical properties arising from the
gravitational forces.

The distribution of galaxies resulting from this GalICS simulation has been
compared with the 2dS \citep{2001MNRAS.328.1039C} and the Sloan Digital Sky
Survey \citep{2001misk.conf..249S} and found to be realistic on the angular
scales of $3' \lesssim \theta \lesssim 30'$, see \cite{2006MNRAS.369.1009B}.
The discrepancy in the spatial correlation function for other values of
$\theta$ can be explained by the limits of the numerical simulation. Obviously,
any information on scales larger than the size of the box ($\sim 45$') is not reliable.
The model has also proven to give sensible results for Lyman break galaxies at $z=3$
\citep{2004MNRAS.352..571B}.
It is also possible to model active galactic nuclei \citep{2005MNRAS.364..407C}.

Since it is possible to reproduce reasonable
correlations from semi-analytic modelling of galaxy formation within
this simulation at $z=0-3$, we now attempt to do so also for somewhat higher redshifts.

\subsection{Clumping of dark matter}
We proceed to calculate the clumping factor $C(z)$.
The inhomogeneities of the DM distribution can be calculated using the 
relative clumping of dark matter halos: $\bar\rho_i = \rho_i/\rho_{mean}$,
where $\rho_{mean}$ is the mean density of the dark matter in the universe
and $\rho_i$ is the mean density of DM halo $i$. 

As matter contracts, the density increases, but since the gamma ray emitting volume
also decreases, the net effect is a linear enhancement from the quadratic dependence
on the density. This means that the DM halos will emit as:
\begin{equation}
	\frac {I_{halos}}{I_0} = \frac{\sum_{i}m_i\bar\rho_i}{\sum_i m_i}\cdot C_{halo},
\end{equation}
where $I_0$ is the intensity for a homogeneous universe and
the summation is done over all DM halos and thus $\sum_i m_i = m_{halos}$.
The factor $C_{halo}$ accounts for the modification from the form and
properties of the halo itself. A simple conic DM distribution would give
$C_{halo}=1.6$. The more realistic distribution $\rho(r)=\rho_0\cdot[(1+r)(1+r^2)]^{-1}$,
where $r$ is the radial coordinate relative to the halo radius, gives $C_{halo}=1.1$.
However, the radiation from within the denser part of the halo will also be
subject to more absorption, and so for the sake of simplicity we use $C_{halo}=1$.
We notice that the average relative density over all the halos in the simulation
 is fairly constant, $\left< \bar\rho_i \right> \sim 70$ for $z<5$.

\begin{figure}[here!] 
        \resizebox{\hsize}{!}{\includegraphics{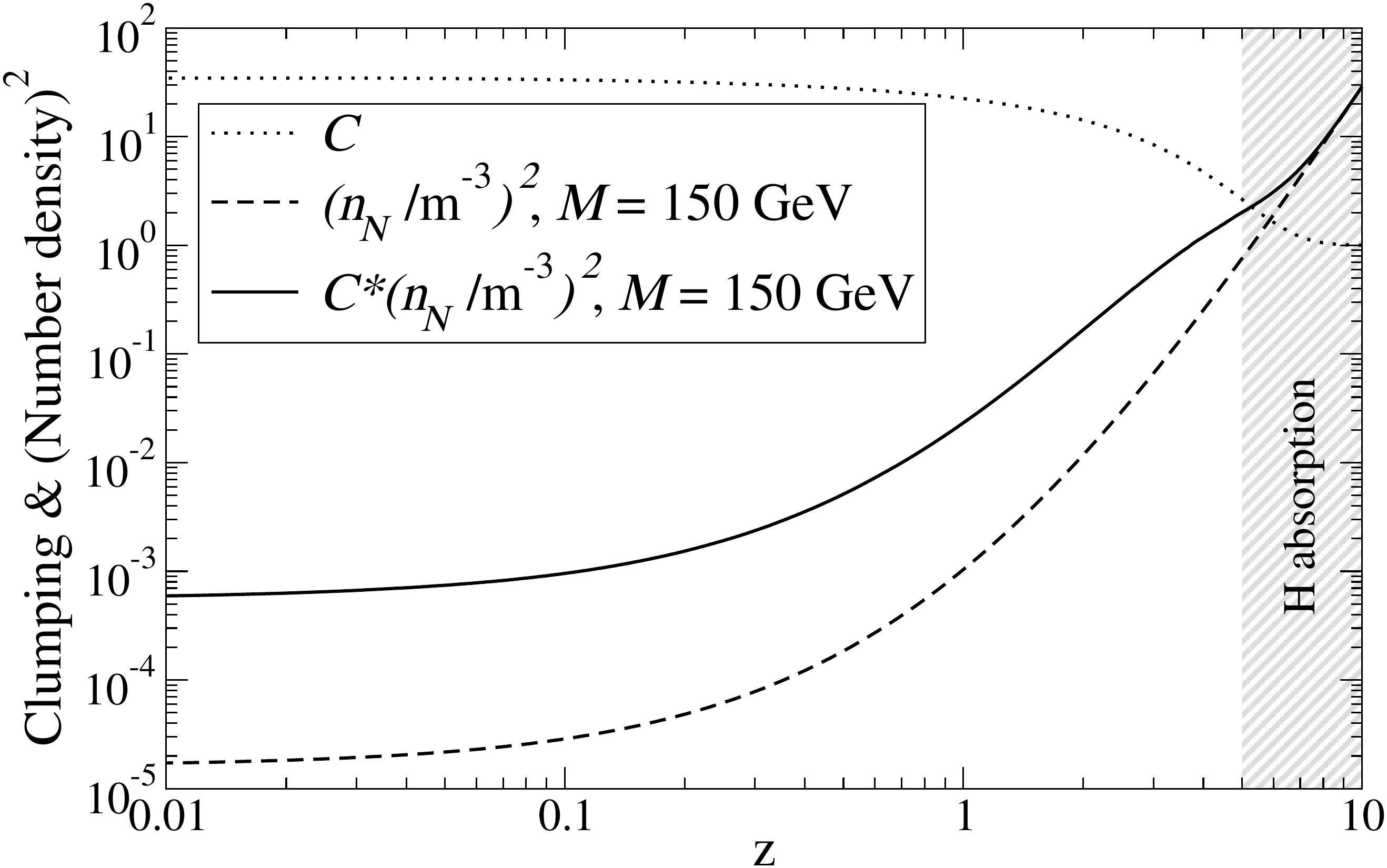}}
        \caption{ The clumping factor ($C$, dotted line) compared to the competing effect of the
	decreasing heavy neutrino number density squared ($n_N^2$, dashed line) for $M_N=150$ GeV
	and the product of the two (solid line). Different neutrino masses scale as in Fig.~\ref{fig:n_T_N}.}
\label{fig:nClump}
\end{figure}

Simultaneously, the DM background (the DM particles that are not in halos) will decrease,
both in density by a factor \linebreak $(m_{tot}-m_{halos})/m_{tot}$ and because of their decreasing fraction
of the total mass in the box $m_{tot}$:
\begin{equation}
	\frac {I_{DM-background}}{I_0} = \left( \frac{m_{tot}-m_{halos}}{m_{tot}}\right)^2.
\end{equation}
This means that the total clumping factor is
\begin{equation}
	C = \frac {I_{halos}}{I_0} + \frac {I_{DM-background}}{I_0} =
	\frac{\sum_{i}m_i\bar\rho_i}{m_{tot}} + \left( \frac{m_{tot}-m_{halos}}{m_{tot}}\right)^2,
\end{equation}
where the first term starts as unity whereafter it decreases and quickly becomes negligeable
with respect to the second term, which starts at zero, but then rapidly increases.
The total clumping is plotted in Fig.~\ref{fig:nClump} along with the competing $(n_N/\textrm{m}^{-3})^2$ effect, as
well as the product, all as a function of the redshift $z$. The number density of heavy neutrinos
in the figure
is taken for the mass $M_N=150$ GeV. 
We notice that the clumping enhancement remains $\sim 30$ for $z<1$ and that the clumping is
$\sim 1$ for $z>5$. This is mainly due to the proportion of mass within the halos
compared to the total DM mass. The clumping enhancement lies between
the two extreme values by \cite{2006PhRvD..73f3504B} and \cite{2005Natur.433..389D} quoted above.

In fact, the clumping factor can be even higher if other halo shapes are assumed
with smaller radii \citep{2002PhRvD..66l3502U}. The densities in the halos considered
in the present work have been evaluated at the virial radius.

We also point out that before the reionization, at $z\gtrsim 5$, there is absorption from neutral hydrogen in
the interstellar medium (ISM), also known as the Gunn-Petersen effect \citep{1965ApJ...142.1633G}.
This means that photons from higher redshifts will be highly
attenuated. For $z=5.3$, the emission drops by roughly a factor of 10, and for $z\sim 6$
the opacity is $\tau_{eff}>20$ \citep{2001AJ....122.2850B}. Hence, any gamma ray signal
prior to this epoch would have been absorbed.


\section{Photon distribution from $N\bar N$-collisions}
In order to evaluate the photon spectrum from $N\bar N$-collisions
we use PYTHIA version 6.410 \citep{2006JHEP...05..026S}. According to \citet[Eq.~13]{1989NuPhB.317..647E}
the centre of mass energy squared is $E_{CM}^2 = 4M_N^2 + 6M_N T_f$ and $T_f \approx M_N/30$ as estimated above.

We generate 100,000 $N\bar N$ events for each mass $M_N=50,60,...,1000$ GeV
and calculate the photon spectrum and mean photon multiplicity and energy. We assume that
$N\bar N$ collisions at these energies and masses can be approximated by
$\nu_\tau\bar\nu_\tau$ collisions at the same $E_{CM}^2$. This is obviously not equivalent, but
$N\bar N$ cannot be directly simulated in PYTHIA. Nevertheless, with the approximations
used in calculating $\left<\sigma v\right>$, the only difference between
$\nu_\tau\bar\nu_\tau$ and $N\bar N$ collisions (except in the cross section)
is the $t$-channel production of $W^+W^-$ through $\tau$. However, since
the heavy neutrinos are non-relativistic when they collide, the two $W$s
will be produced back-to-back, which means that the inclusion of the $t$-channel
is unimportant.

In order to verify this, we study the difference in the photon spectrum for
$W$ decay at 0 and 90 degrees, and despite an increasing difference between
the two cases, even at $M_N=1000$~GeV, the difference is not strong enough to change our conclusions.

\begin{figure}[here!] 
        \resizebox{\hsize}{!}{\includegraphics{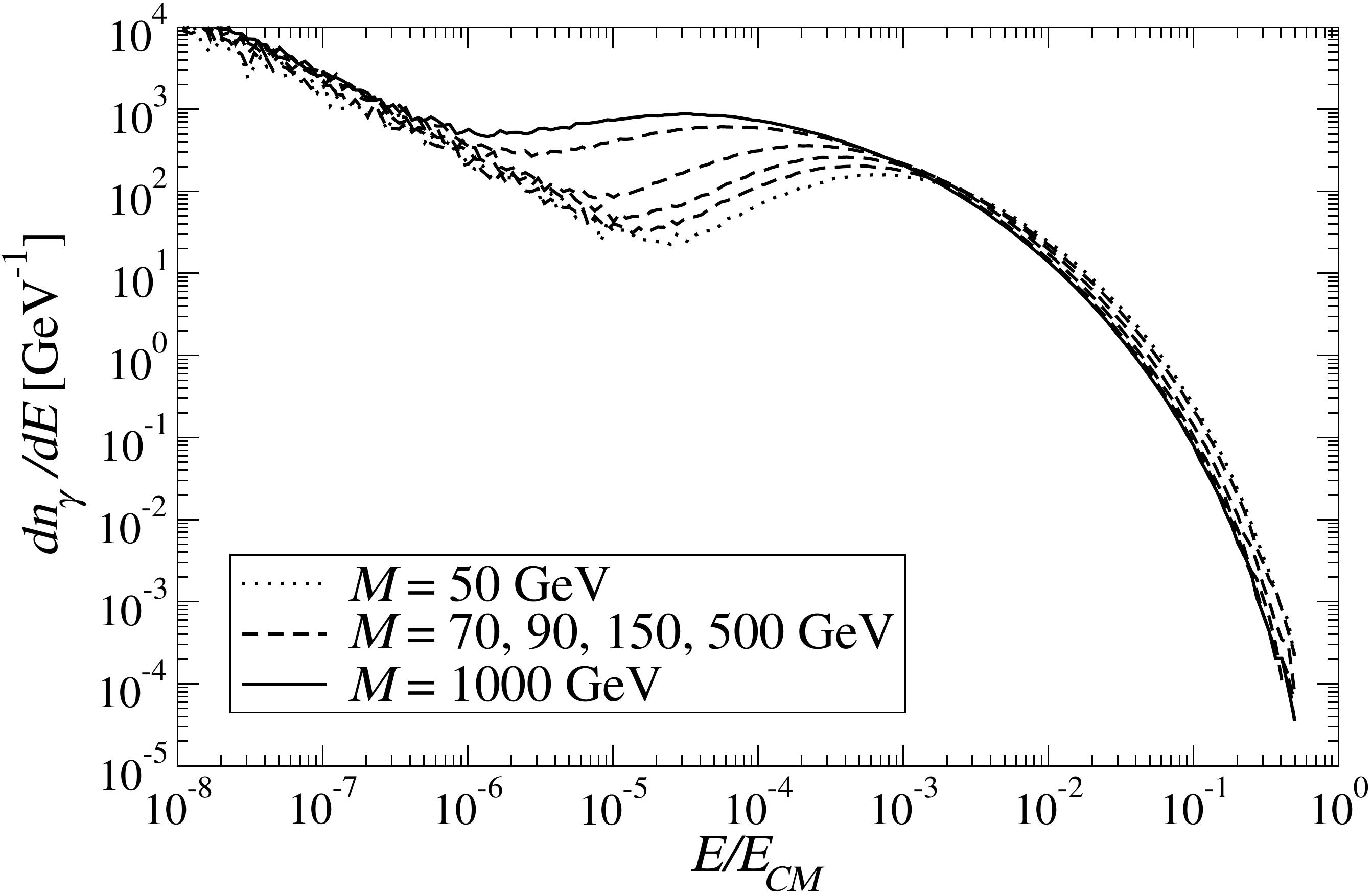}}
        \caption{ The relative energy distributions of photons from $N\bar N$-collisions for
	heavy neutrino masses $M_N=$ 50, 70, 90, 150, 500, 1000 GeV. $E_{CM}=2M_N$ is the centre of mass energy.}
\label{fig:dndE}
\end{figure}

\begin{figure}[here!] 
        \resizebox{\hsize}{!}{\includegraphics{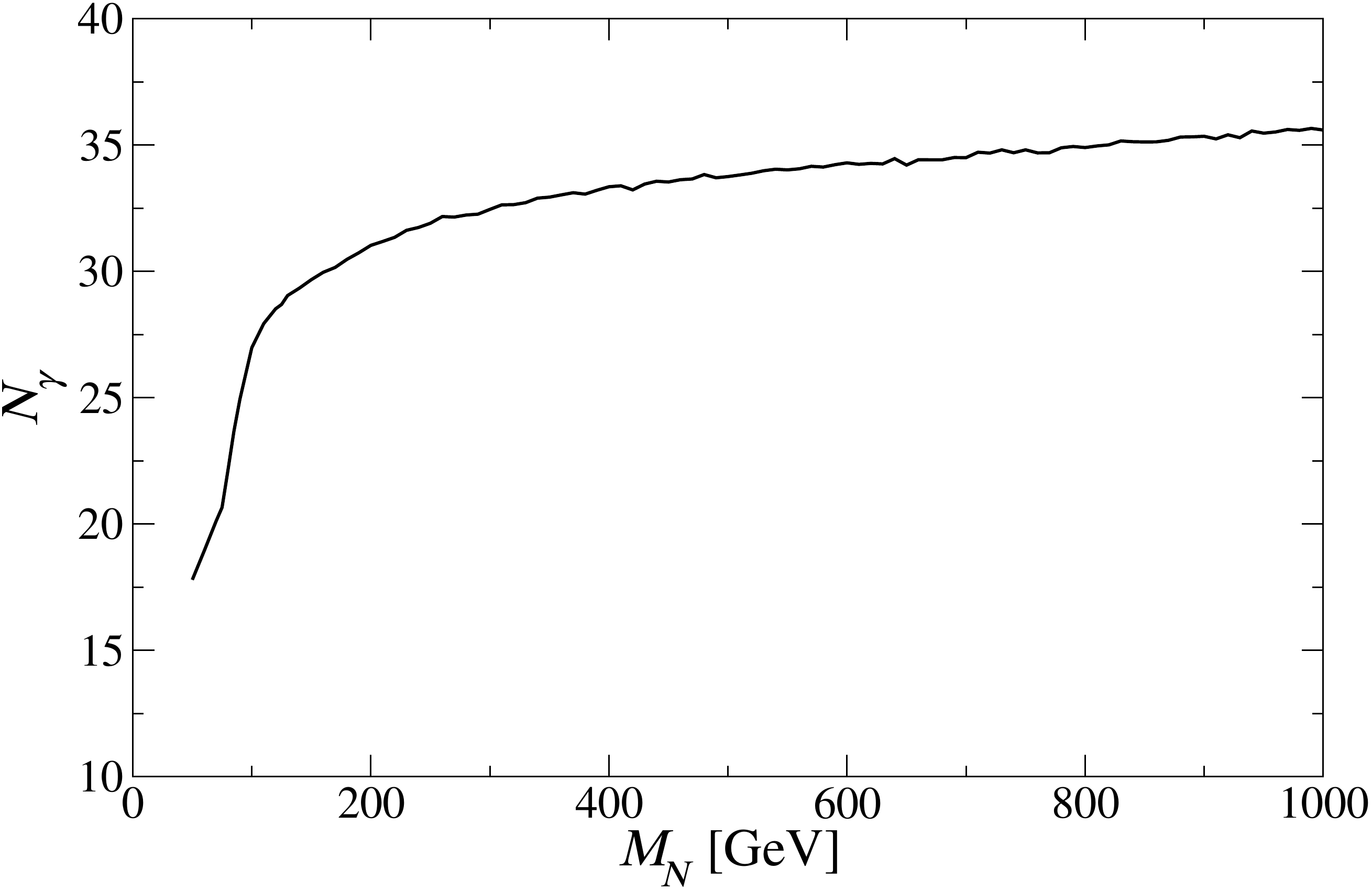}}
        \caption{ The average number of photons produced for an $N\bar N$-collision as a function of
	heavy neutrino mass $M_N$ in GeV.}
\label{fig:nTot_M}
\end{figure}

The resulting photon distribution is presented in Fig.~\ref{fig:dndE}.
We note that the photon energies peak at $E_{CM}/2$, which is natural
since the decaying particles can each have at most half of the centre of mass energy.
The curves continue to increase as $\propto E^{-1}$ as $E$ decreases further.
Note that the noise in the curves for lower $E$ is due to lacking statistics for these
rare events, but it does not affect the outcome of the calculations.
We also calculate the mean photon energy and find it to be $\bar E_\gamma \approx 0.21 E_{CM}$
for all masses. 
The curve is normalized such that the integral over $ \frac{dn_\gamma}{dE} $ is unity.
The average number of photons, $N_\gamma$, produced for an $N\bar N$-collision is shown in Fig.~\ref{fig:nTot_M}.
The sharp rise in the curve at $M_N\sim100$ GeV is due to the jets from the emerging $W^+W^-$-production.

\section{Gamma ray spectrum}
The $N\bar N$-collisions from the reionization at $z_i\sim 5$ until today
give an integrated, somewhat redshifted, gamma spectrum for a heavy neutrino
with a given mass:
\begin{equation}
	I =
	\int_{T_i}^{T_0} 
C(T) \frac{n^2\left<\sigma v\right> }{4\pi} N_\gamma \frac{dn_\gamma}{dE}\left|_{E\frac{T_0}{T}}\right.\frac{dt}{dT} dT,
\end{equation}
where
$C(T)$ is the clumping factor in Fig.~\ref{fig:nClump} and
$\frac{dn_\gamma}{dE}$ is the photon distribution in Fig.~\ref{fig:dndE}.
$T_0=2.725$ K is the temperature of the CMB today and $T_i$ is the reionization temperature,
which we set to $T_i = 5\cdot T_0$.

\begin{figure}[here!] 
        \resizebox{\hsize}{!}{\includegraphics{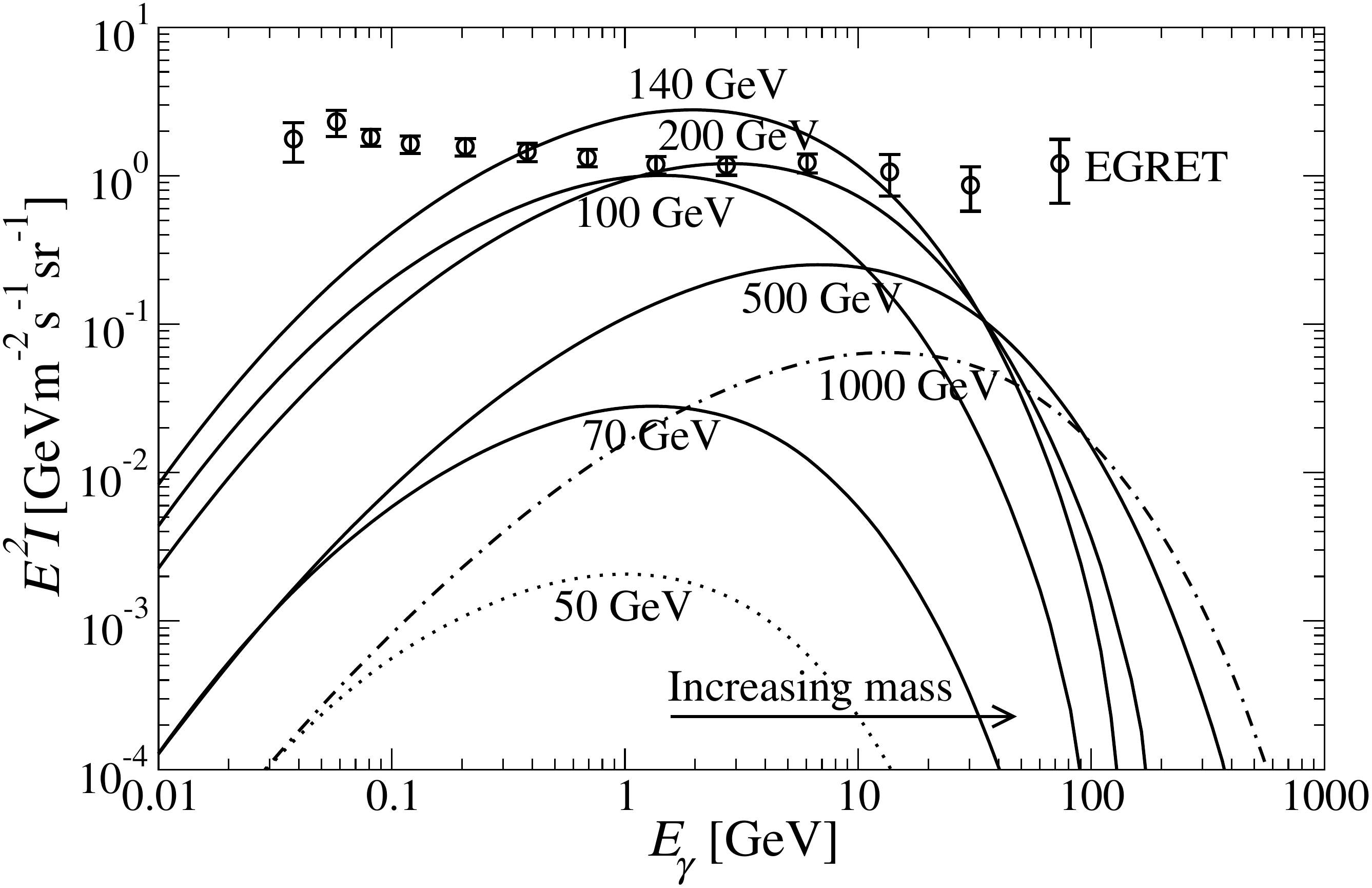}}
        \caption{ Cosmic gamma radiation from photons produced in $N\bar N$-collisions
	as a function of photon energy for neutrino masses $M_N=50, 70, 100, 140, 200, 500, 1000$ GeV.
	The dotted line represents $M_N=50$ GeV
	and the dot-dashed $M_N=1$ TeV.
	The solid lines are the masses in between. The circles represent data from EGRET \citep{1998ApJ...494..523S},
	with error bar, as derived for extragalactic sources.}
\label{fig:EEI}
\end{figure}

\begin{figure}[here!] 
        \resizebox{\hsize}{!}{\includegraphics{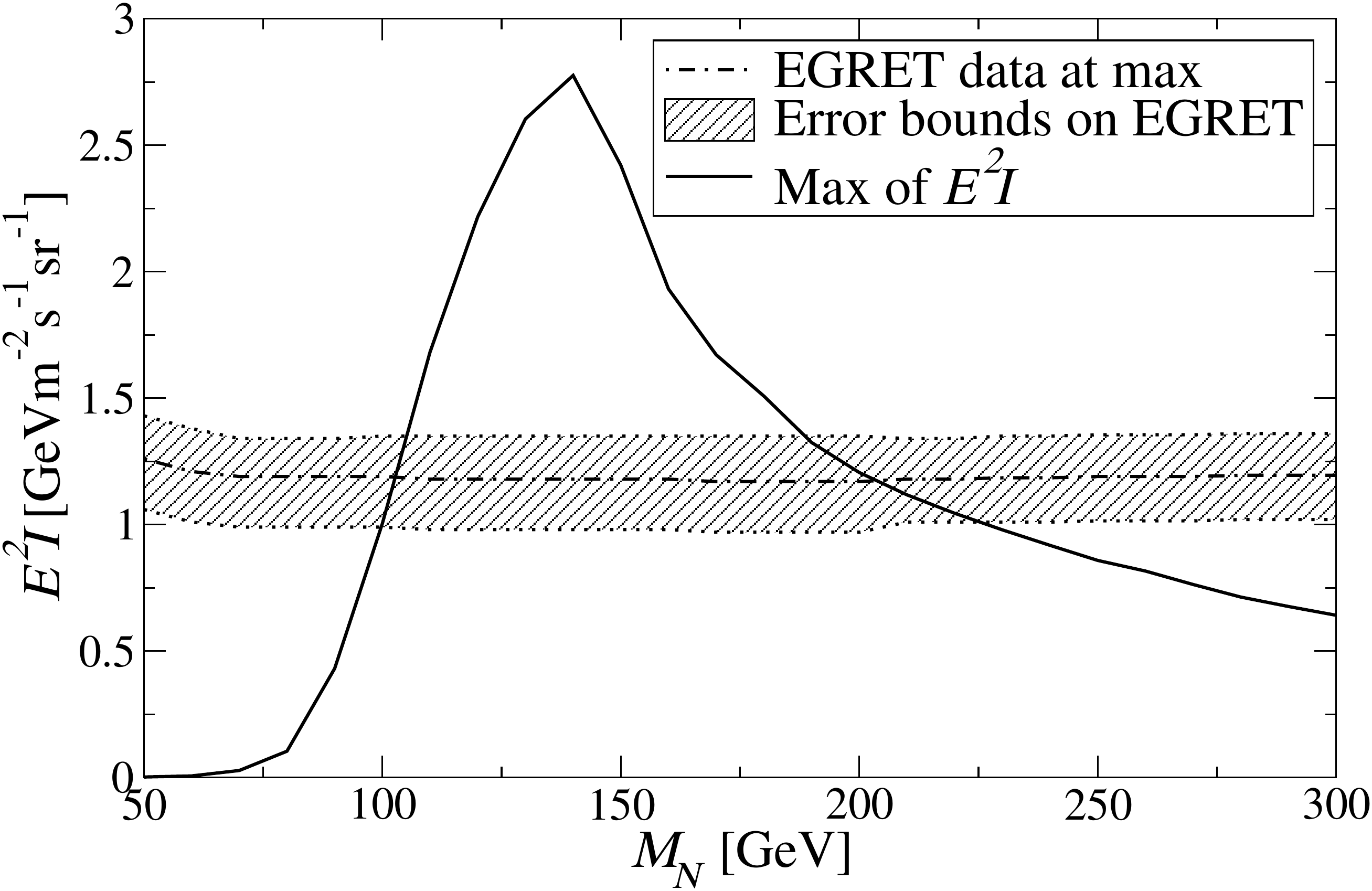}}
        \caption{ Maximum cosmic gamma radiation from photons produced in $N\bar N$-collisions
	as a function of neutrino mass (in GeV).
	The marked region is excluded since $\Omega_N>\Omega_{DM}$ within.
	The data are taken at the energy corresponding to the maximum in Fig.~\ref{fig:EEI}.
	with error bars.}
\label{fig:EEImax}
\end{figure}

The resulting $E^2I$ is presented in Fig.~\ref{fig:EEI}. When we compare the calculated heavy neutrino signal
with data from EGRET \citep{1998ApJ...494..523S}, we see that only neutrino masses around $M_N \sim 100$ or 200 GeV
would be detectable, and then only as a small bump in the data around $E_\gamma\sim 1$ GeV. 
For intermediary neutrino masses, the signal would exceed the observed gamma ray data.
In Fig.~\ref{fig:EEImax}, the peak intensity for the different heavy
neutrino masses is plotted, as well as EGRET data for the corresponding energy with error bars. 
The data 
represent the observed diffuse emission at high latitudes ($\left|b\right|>10$ degrees),
where first the known point sources were removed and then the diffuse emission in our galaxy was subtracted.

We have also compared the height of the curves, both with and without clumping,
and the integrated difference is roughly a factor of 30.

\section{Discussion and conclusions}
\label{sec:Discussion}

The numerical calculation of the evolution of the heavy neutrino number density
indicates that in the mass region $100\lesssim M_N \lesssim 200$, the cosmological neutrinos would
give a cosmic ray signal that exceeds the measurements by the EGRET telescope \citep{1998ApJ...494..523S}.
Note that the clumping factor for these limits is rather conservative.
In \citet{2002PhRvD..66l3502U}, this factor is much larger, which would also
produce a stronger limit on the heavy neutrino mass.

\begin{figure}[here!] 
        \resizebox{\hsize}{!}{\includegraphics{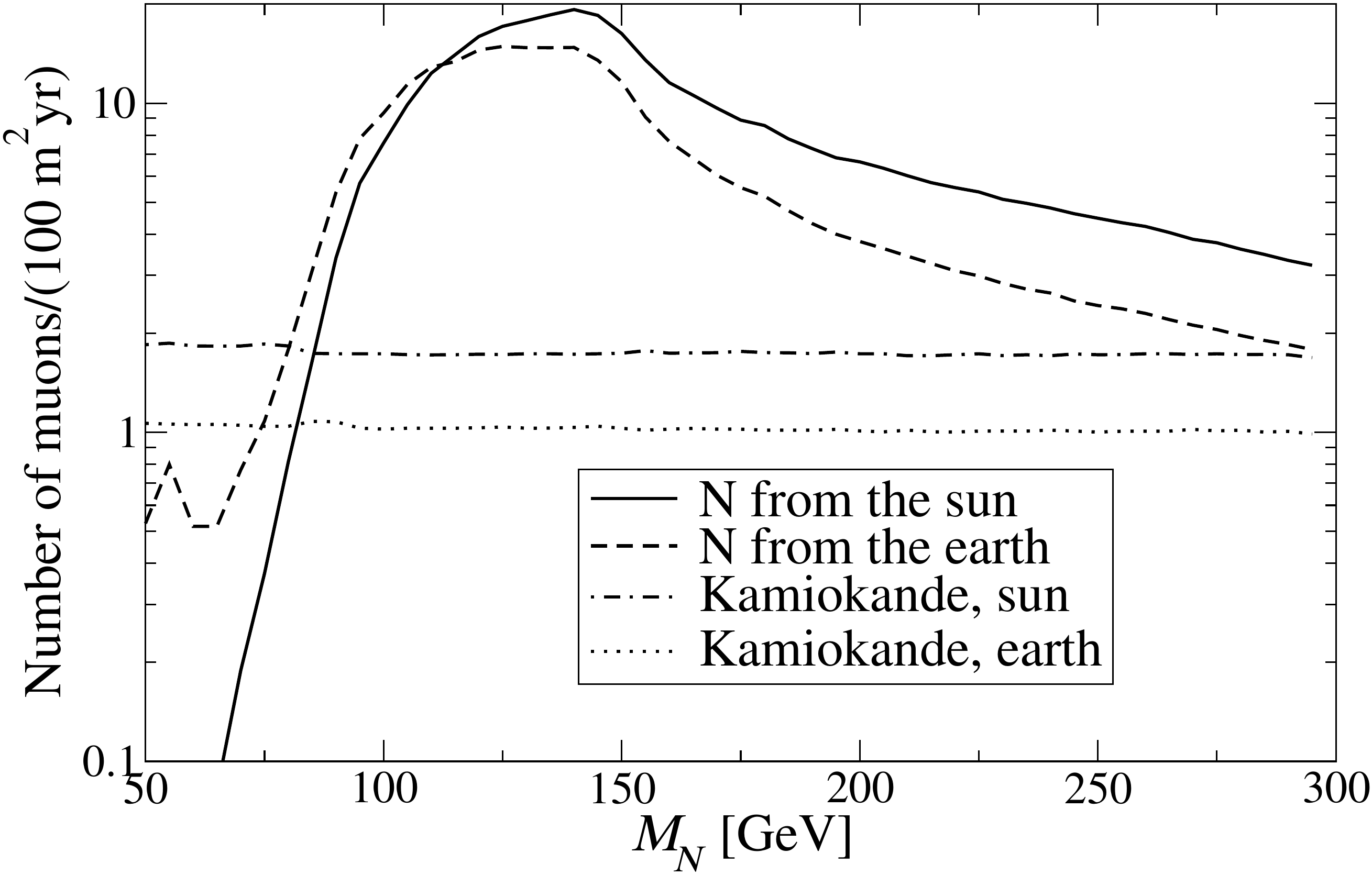}}
        \caption{
	Predicted signal from enhanced $N\bar N$ annihilation in the earth and the sun
	compared to the measured signal in the Kamiokande.
	On the y-axis: the number of muons (per 100~m$^2$year) produced by muon neutrinos resulting from heavy neutrino collisions
	in the sun and the earth, as evaluated by \cite{1992PhLB..289..463M}, but scaled to
	our $\Omega_N(M_N)$. On the x-axis: the heavy neutrino mass in GeV.
}
\label{fig:compKamiokande}
\end{figure}

We can also compare our neutrino density with the results from the Kamiokande collaboration
\citep{1992PhLB..289..463M}. We scale the neutrino signal in their Fig.~2 to $\Omega_N/\Omega_{DM}$,
where we use $h_0=0.71$, $\Omega_{m}=0.2678$ and $\Omega_b=0.044$. This is shown in Fig.~\ref{fig:compKamiokande},
where we compare our numerical results for the relic neutrino density to the observed muon flux
in the Kamiokande detector.
This gives an exclusion region of $80\lesssim M_N \lesssim 400$ GeV.
Our analytical results, which are comparable to the traditional
relic neutrino densities, is about a factor two lower, giving an exclusion region of $90\lesssim M_N \lesssim 300$~GeV.
The model that gives these limits \citep{1987ApJ...321..571G} is rather
complicated and not verified experimentally, so these results cannot be taken strictly.
Note also that in the three-year WMAP analysis \citep{2007ApJS..170..377S}, the value of $\Omega_{DM}$
depends on which other data the WMAP data are combined with. For WMAP+CFHTLS $\Omega_{DM}$
can be as high as 0.279 and for WMAP+CBI+VSA it can be as low as 0.155.
The higher of these possibilities would give an exclusion region of $85\lesssim M_N \lesssim 350$ GeV.
The lower boundary value would give an exclusion region of $75\lesssim M_N \lesssim 500$ GeV.
A conservative limit based on the Kamiokande data gives the exclusion region $100\lesssim M_N \lesssim 200$ GeV.


If a heavy neutrino exists with a mass $M_N\sim100$ GeV or $M_N\sim200$ GeV
it would give a small bump in the data at $E_\gamma\sim1$ GeV.
Currently the
data points are too far apart and the error bars too large to neither exclude
nor confirm the eventual existence of such a heavy neutrino.
Most of this part of the gamma ray spectrum is usually attributed to blazars,
which have the right spectral index, $\sim 2$ \citep{1997ApJ...490..116M}.



We note that there could be an enhancement in the signal due to the higher DM densities
within galaxies compared to the mean density in the halos. On the other hand, from within
galaxies there will also be an attenuation due to neutral hydrogen, thus reducing
the enhancement. There will also be a certain degree of extinction of the signal
due to neutral hydrogen along the line of sight, but even if we assume complete extinction
above $z=4$ the resulting spectrum decreases with only about 20\%.



We are also aware of the ongoing debate concerning the antiprotons -- whether or not the
DM interpretation of the EGRET gamma excess is compatible with antiproton measurements
\citep{2006JCAP...05..006B, 2006astro.ph.12462D}.
We note the argument by de Boer that antiprotons are sensitive to electromagnetic fields,
and hence their flux need not be directly related to that of the photons, even if they
too were produced by $N\bar N$ annihilation.

In the advent of the Large Hadron Collider, we also point out that there may be a possibility
to detect the existence of a heavy neutrino indirectly through the invisible Higgs boson
decay into heavy neutrinos 
\citep{2003PhRvD..68e4027B}.


It will of course be interesting to see the results of the gamma ray large area space telescope (GLAST).
It has a field of view about twice as wide (more than 2.5 steradians),
and sensitivity about 50 times that of EGRET at 100~MeV and even more at higher
energies. Its two-year limit for source detection in an all-sky survey is $1.6 \times 10^{-9}$
photons cm$^{-2}$~s$^{-1}$ (at energies $>$ 100~MeV). It will be able to locate sources
to positional accuracies of 30 arc seconds to 5 arc minutes.
The precision of this instrument could well be enough to detect a heavy neutrino signal
in the form of a small bump at $E\sim 1$ GeV in the gamma spectrum,
if a heavy neutrino with mass $\sim$100 or 200~GeV would exist.

There are also some other possible consequences of heavy neutrinos that may be worth investigating.
The DM simulations could be used to estimate the spatial correlations that the gamma rays would have
and to calculate a power spectrum for the heavy neutrinos. This could be interesting at least for masses
$M_N\sim 100$~GeV and $M_N\sim 200$~GeV. The annihilation of the heavy neutrinos could also help to explain
the reionization of the universe.
Another possible interesting application of heavy neutrinos would be the
large look-back time they provide
\citep{2006PhLB..639...14S}, with a decoupling temperature of $\gtrsim 10^{13}$ K \citep{1989NuPhB.317..647E}.


\begin{acknowledgements}
E. E. would like to express his gratitude to
Konstantine Belotsky,
Lars Bergstr\"om,
Michael Bradley,
Alexander Dolgov,
Kari Enqvist,
Kimmo Kainulainen
and
Torbj\"orn Sj\"ostrand
for useful discussions and helpful comments and explanations.
We are both grateful to the GalICS group, who has provided the complete dark
matter simulations and finally to the Swedish National Graduate School
in Space Technology for financial contributions.
\end{acknowledgements}



\bibliographystyle{aa} 


\bibliography{bibtex}        

\end{document}